\documentclass[a4paper,11pt]{article}
\usepackage{pos}
\usepackage{float}
\usepackage{multirow}
\usepackage{wrapfig}
\usepackage{enumitem}
\usepackage{graphicx}

\title{Simultaneous observations of multiple ELVES and SPRITES at  the Pierre
Auger Observatory}
\ShortTitle{Investigating ELVES, SPRITES and halos with
the Pierre Auger FDs}

\author*[a]{Roberto Mussa}

\onbehalf{for the Pierre Auger Collaboration$^b$}

\affiliation[a]{Istituto Nazionale di Fisica Nucleare, Sezione di Torino, via Pietro Giuria 1, 10125, Torino, Italia}

\affiliation[b]{Observatorio Pierre Auger, Av.\ San Mart{\'\i}n Norte 304, 5613 Malarg\"ue, Argentina\\
Full author list: \normalfont{\url{https://www.auger.org/archive/authors\_icrc\_2025.html}}}

\emailAdd{spokespersons@auger.org}

\abstract{Since 2014, the Pierre Auger Observatory has exploited a dedicated trigger and its very high time resolution to study ELVES and harvest record samples of multiple ELVES using the Fluorescence Detector (FD). In 2017, after extending the readout of trace lengths to 0.9 ms, we started observing other types of light transients from the base of the ionosphere, such as HALOS, which deserved further investigation. In December 2023 and April 2024, we installed two additional cameras (TLECAMs), which allow us to perform simultaneous detection of these transients with higher space resolution and longer integration times. Here, we present our first simultaneous observations of SPRITES and ELVES by both TLECAMs and FD. Furthermore, we describe the Python algorithm based on DBSCAN to automatically detect SPRITES in the videos recorded by our TLECAMs and acquire data efficiently without needing the FD trigger.}

\FullConference{
39th International Cosmic Ray Conference (ICRC2025)\\
15-24 July, 2025\\
Geneva, Switzerland
  }

\begin{document}
\maketitle

\section{Introduction}
The Pierre Auger Observatory is the world's largest infrastructure for the study of ultra-high energy cosmic rays. Besides its main activity, the Observatory has started a program of cosmo-geophysics studies, which exploit some of the unique features of its detectors. ELVES (Emission of Light and Very low-frequency perturbations due to Electromagnetic pulse Sources) are transient luminous events occurring at the base of the ionosphere when a strong electromagnetic pulse (EMP) is emitted  by lightning. This phenomenon, theoretically predicted a few years before \cite{Inan:1991}, was photographed for the first time in 1990 from the Space Shuttle \cite{Boeck:1992}. For an observer on the ground, ELVES appear as rapidly expanding rings, smoothly fading towards the horizon, and can be observed at distances up to more than 1000 km from  our  fluorescence detector (FD) \cite{Abraham:2009pm}.  The center of the expanding rings is not above the vertical of the causative lightning but at the point of the minimum light path between source and observer, as we explained in our previous papers \cite{Mussa:2020vap,PierreAuger:2021ecs}.

\begin{figure}[h]
\centering
    \includegraphics[height=7cm]{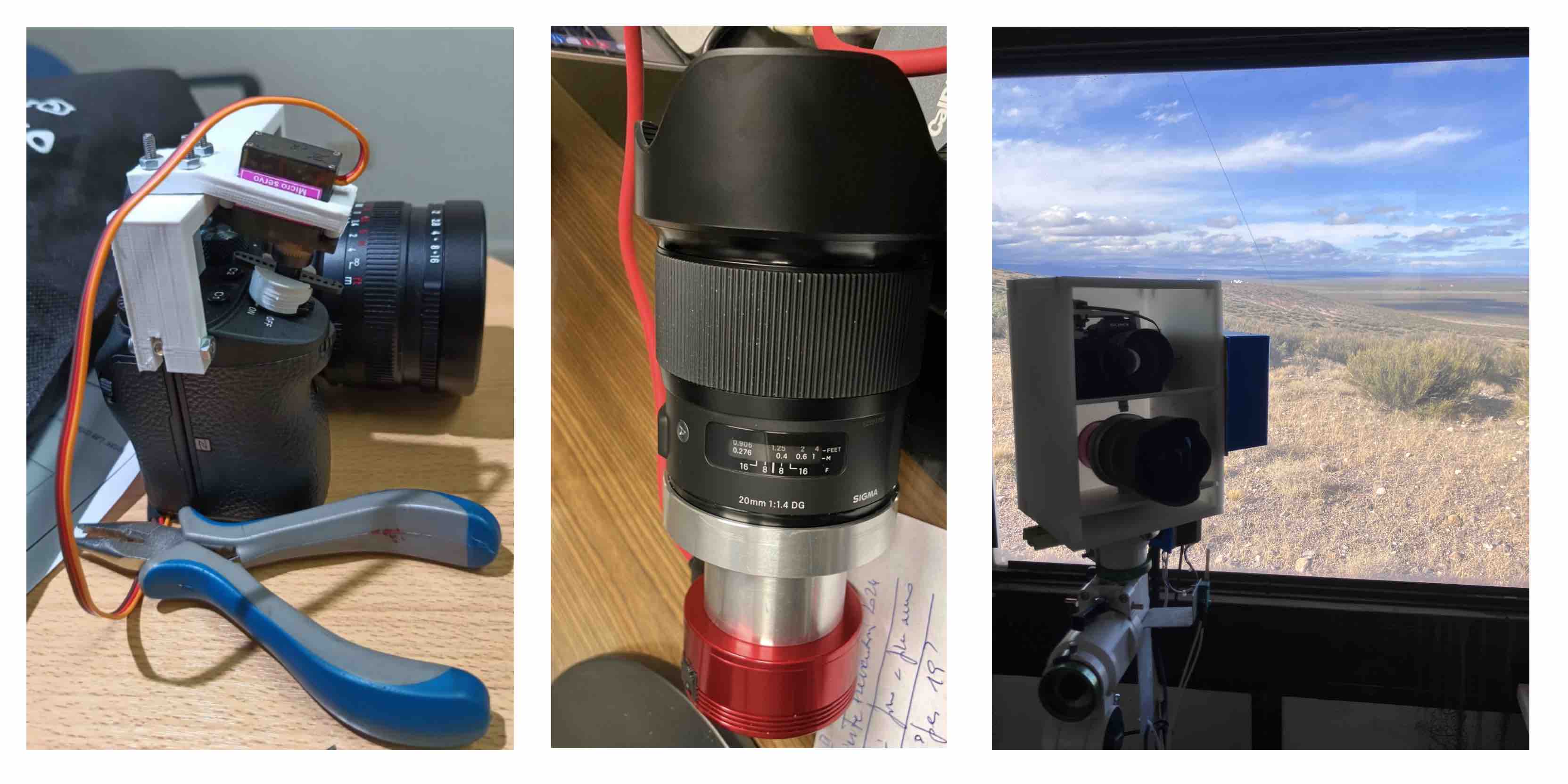} 
    \caption{The new TLECAMs: Sony $\alpha$7-III (left), ZWO ASI294MC (center), mounted in parking position with the field of view on the back (right). }
    \label{fig:TLEcams}
\end{figure}

The events triggered by the Auger Observatory are produced by lightning sources  far  enough from the FD, i.e. more than 250 km away, so that the earth's curvature prevents the direct light   from reaching our sensors. In 2013, we developed a dedicated trigger and readout scheme to study these events using the twenty-four FD Telescopes \cite{Tonachini:2011zz,Mussa:2012dq,ThePierreAuger:2013eja,PierreAuger:2020lri}. Since 2017, we have extended the length of recorded traces to up to 0.9 ms to measure the doughnut-shaped region of maximum emission fully. Such an upgrade has opened perspectives for observation of other types of Transient Luminous Events (TLEs from now on), such as the halos, as we reported at the previous ICRC conference \cite{PierreAuger:2023gmv}. In December 2020, we implemented the ELVES trigger and readout in the data-taking of the three additional high-elevation telescopes (HEAT \cite{Mathes:2011zz}), to allow observing ELVES from closer lightning at distances between 150 km and 250 km.

\begin{table}
\centering
\caption{Technical specifications of TLEcam-1 and TLEcam-2}
\label{tab:TLEcams}
\begin{tabular}{|llll|}
\hline
&  camera type & sensor size & objective \\
\hline
TLEcam-1 & Sony $\alpha$7-III &35.8x23.8 mm$^2$ (6000x4000)  & 7artisans 50mm f/0.95 \\
TLEcam-2 & ZWO ASI294MC & 19.1x13 mm$^2$ (4144x2822) & Sigma 20mm f/1.4 \\
\hline
\end{tabular}

\end{table}

\section{Complementing Auger FD observations with TLE cameras}

The main drawback of the high time resolution of the FD is the limit on the recordable trace length, which cannot exceed 1 ms without introducing large dead time on the DAQ for cosmic ray observations (a 0.9 ms event requires 9 seconds to be written to disk). A second limitation comes from the memory depth of the second-level trigger, which can take up to 64 consecutive traces, i.e. 6.4 milliseconds. In a significant fraction of ELVES triggers, we observed full saturation of the second-level memory buffers, suggesting extended emission of light from the horizon. In most cases, despite the top of the thunderstorm cloud being below the horizon, multiple scattered light from the bolt can reach our sensors a few ms after the ELVES.

\begin{figure}[h]
\centering
    \includegraphics[height=7cm]{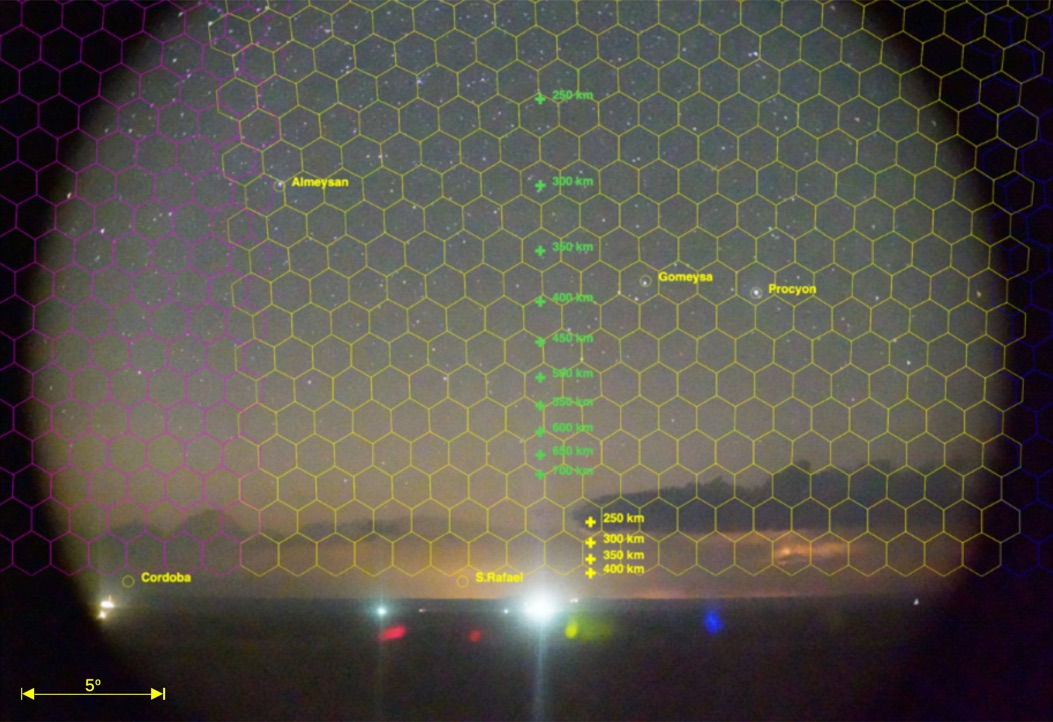} 
    \caption{Sony camera alignment with brightest stars. Green crosses indicate the elevation  of a potential ELVES center (h=90 km) as a function of the distance of the lightning source. Yellow crosses indicate the elevation of the top of a 15 km high cloud as a function of the distance of the lightning source. The colored hexagons show the field of view of each FD pixel.}
    \label{fig:star_align}
\end{figure}

Various other types of “transient luminous events” (TLEs) can indeed occur at the time of the lightning: besides halos, we can expect to detect SPRITEs (the acronym stands for Stratospheric/mesospheric Perturbations Resulting from Intense Thunderstorm Electrification) or blue and gigantic jets. SPRITEs were theoretically predicted by CTR Wilson in 1925 \cite{Wilson:1925} to occur under strong thunderstorms when a positive cloud-to-ground strike is taking place. SPRITEs, recorded for the first time in 1989 \cite{Franz:1990}, are intense discharges moving both upwards and downwards from 70 to 40 km altitude, with widths of about 0.1 km  and time scales of about 5 ms to 100 ms. Such discharges can appear almost simultaneously over a horizontal distance of several km, have peculiar shapes (similar to jellyfish or carrots), and are significantly brighter than the ELVES. Blue and gigantic jets 
\cite{Wescott:1995} last even longer than SPRITEs, start from the top of high thunderstorm clouds, and can last a few tenths of a second, and can reach the base of the ionosphere.
To complement the FD observations of ELVES and halos, we installed, in the proximity of the FD site at Coihueco, two cameras with high photon sensitivity, good space resolution and a field of view comparable with that of each FD telescope, capable of recording video sequences at speeds of up to 100 fps. The specifications of these two cameras are summarised in Table \ref{tab:TLEcams}. Both cameras are read by an Intel NUC PC via USB-3 ports. This paper reports the preliminary results from the first year of commissioning of these new devices. 

\begin{figure}[h]
\centering
    \includegraphics[width=14cm]{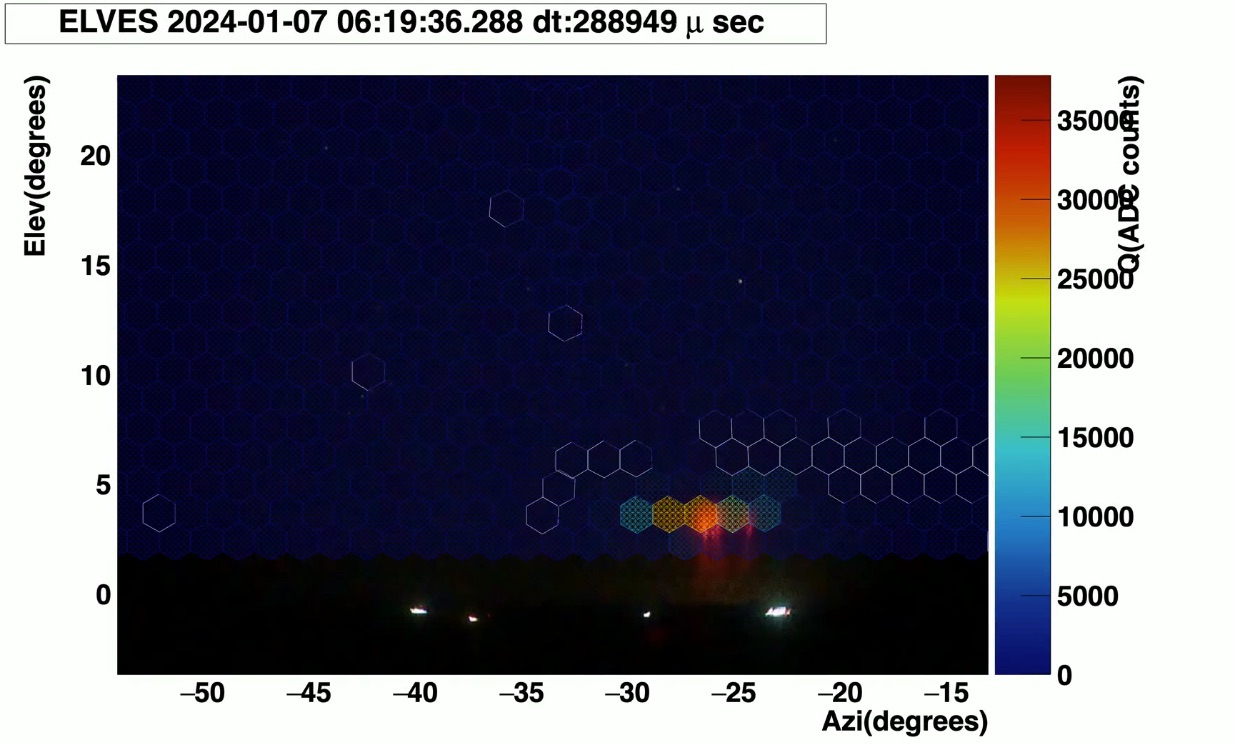}
    \caption{The SPRITE image recorded by TLEcam-1 at GPStime 1388643594 overlapped with the final frame of the ELVES detected by the Coihueco FD.  }
    \label{fig:elve_sprite_1}
\end{figure}

The Sony camera (named TLEcam-1) was installed in December 2023 and could take data during the Austral summer 2023-2024. The ZWO camera
(named TLEcam-2) was installed in April 2024, after the storm season, and the first
observation of a TLE (consisting of an ELVE, followed by a SPRITE) by
both cameras occurred on November 28, 2024. 

Both cameras are mounted in
the same enclosure, shown in Fig.\ref{fig:TLEcams}, which can be remotely controlled via an Arduino microprocessor to rotate azimuthally over a range of about 90 degrees centered in the North-East direction, looking towards the city of Cordoba. 
The Arduino processor is also used to power on and off TLEcam-1 via a custom-made actuator. Synchronization between FD observations and TLE events is, at present, not better than 100 ms. The alignment between the two cameras and the field of view of the FD telescopes has been performed using the brightest stars, as shown in Fig. \ref{fig:star_align}, which also shows the elevation of a ninety-km high ELVES center (green crosses) and  a fifteen-km high cloud top (yellow crosses) as a function of its distance. Throughout the whole commissioning period, cameras were manually steered to follow the storm evolution using the online data browser GeoRayos \cite{GeoRayos:2021}, available at https://georayos.citedef.gob.ar/.

\section{First events observed by TLEcam-1}

TLEcam-1 records 40 ms frames in 5-minute files. The first events were observed in the morning of December 13, 2023: we detected a few ELVES triggers with our FD, and the first four SPRITEs events. Only one SPRITE was correlated with the ELVES within 0.1 s. In principle, we do not necessarily expect a direct time correlation between ELVES and SPRITEs, but  the observation of five SPRITEs in closer time coincidence with ELVES events on January 7, 2024, indicates that the causal connection between these two types of TLE may depend on the type of thunderstorm. 

\begin{figure}[h]
\centering
    \includegraphics[width=6cm]{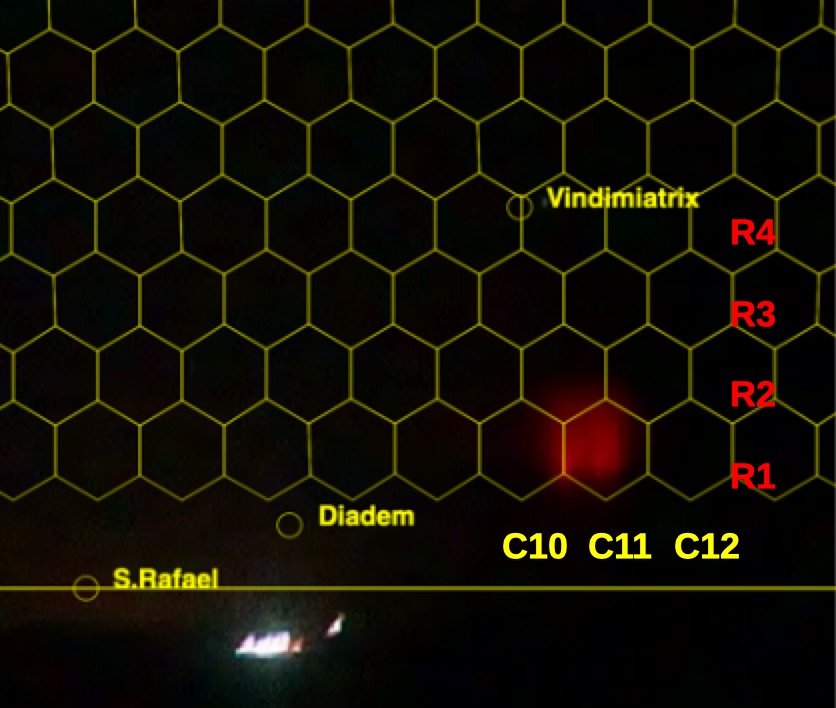}
    \includegraphics[width=9cm]{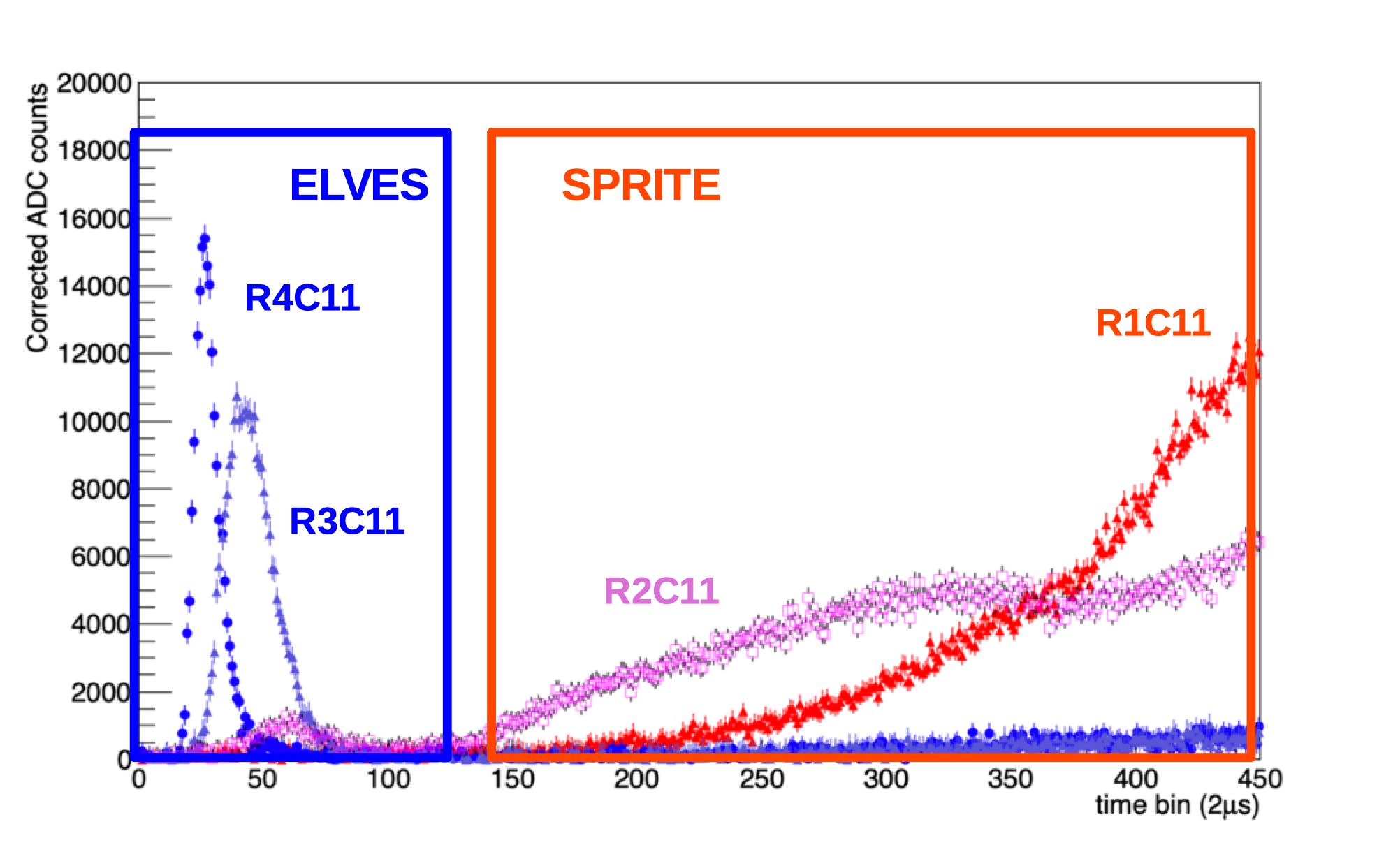} 
    \caption{Left: SPRITE image, overlapped to the pixels grid. Right: the traces corresponding to the indicated pixels, showing the initial part of the SPRITE pulse.}
    \label{fig:elves_traces}
\end{figure}

Fig.\ref{fig:elve_sprite_1} shows one of these events, overlapping the SPRITE profile with the light profile recorded by the FD pixels at the same time, after the ELVES. As this event occurred at an estimated distance of 650 km, the light of the ELVES (much dimmer than the one of the SPRITE) is not visible in the video recorded by the Sony camera. Only one of the five events (recorded at GPS time = 1388641179, shown in 
Fig.\ref{fig:elves_traces}) shows a clear increase in light intensity from the pixels spatially overlapped to the SPRITE image. The left plot shows that the SPRITE is mostly contained in pixel C11R1. The integral of the light curves in pixels C11R3,4 related to the ELVES is one order of magnitude or more smaller than the following light pulse, seen only in pixels C11R1,2. The absence of similar behavior in the other events suggests that in most cases the SPRITE can occur many milliseconds after the ELVES.

\begin{figure}[h]
\centering
    \includegraphics[width=14cm]{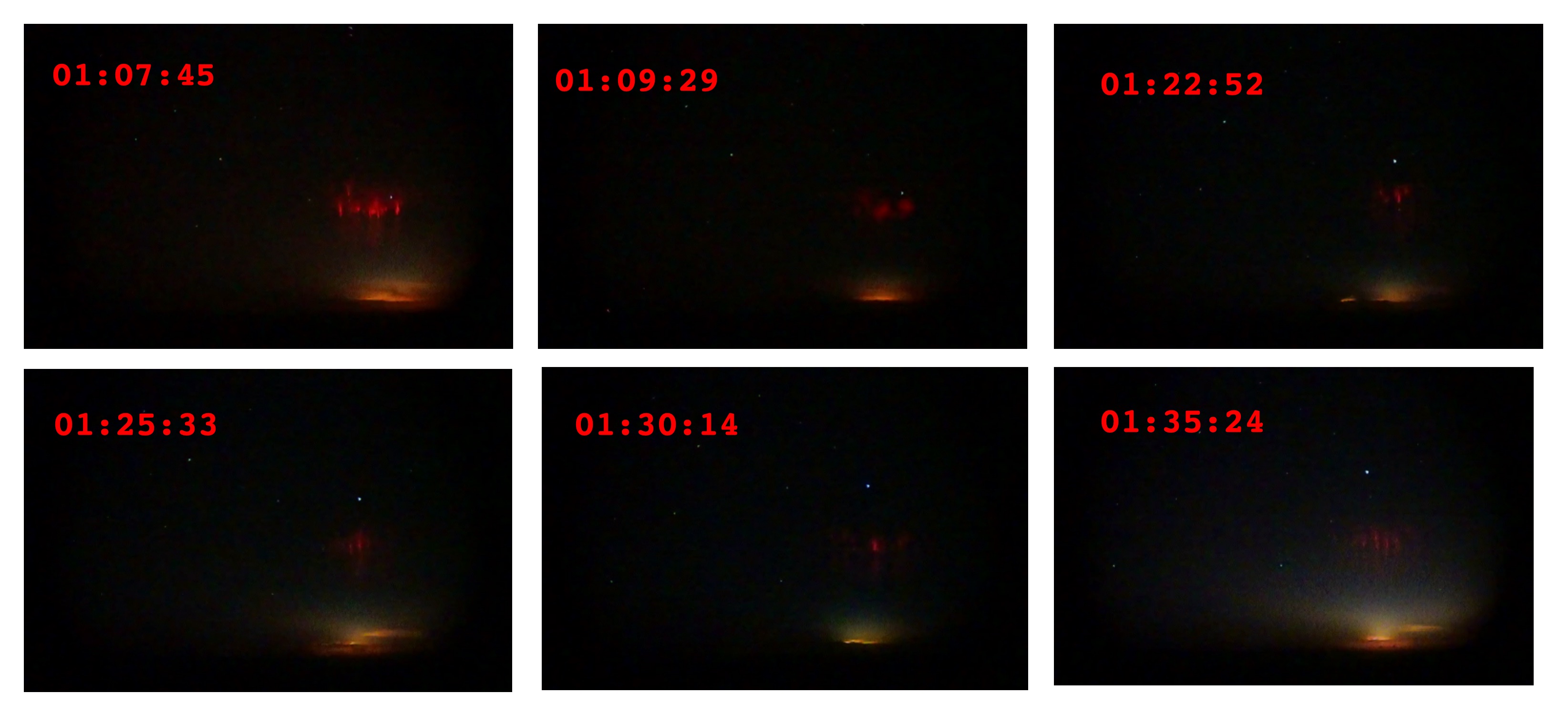}
    \caption{Six SPRITEs recorded in less than half an hour on March 9, 2024.}
    \label{fig:six_sprites}
\end{figure}

Another storm on March 9, 2024, yielded seven SPRITEs observations, none of which had an ELVES trigger in the same GPS second. Six of these events are shown in Fig.\ref{fig:six_sprites}. The glow below the SPRITE is due to Rayleigh scattered light coming from the top of the thunderstorm cloud.

\section{Python code for automatic detection of TLE}

From the first months of commissioning, the observation of a very large number of TLE events without correlation to ELVES triggers urged us to design software code which was capable of processing the large data flow of images produced by TLEcam-2 
(typically 10 GB in 10 minutes), in almost real time, and detect SPRITES by simply
comparing consecutive frames.
Therefore we made use of Python libraries to (a) convert color images to grey-scale (PIL.Image.convert), and (b) finally to numerical arrays (NUMPY.array), (c) to create 2D-vectors above a certain threshold (NUMPY.argwhere), and (d) to search for clusters of points in the resulting arrays (SKlearn.DBSCAN).
This allows for reducing the size of recorded data by two or more orders of magnitude
 while keeping only about sixteen frames in the time window of about one second in close proximity with the SPRITE candidate. When the storm is close to the Observatory, the most dominant residual background is due to the Rayleigh scattered light from the lightning and can be significantly reduced by properly adjusting the lower edge of the analysed image. 
After this cut, few background events due to meteors, trucks’ headlights, and bright stars remain. Processing times ($\approx$ 15 minutes to process 10 minutes of data) are not sufficient to filter the data in real time, but allow for processing them during the following day.

\section{First TLE events observed by both cameras}

\begin{figure}[h]
\centering
    \includegraphics[height=4.5cm]{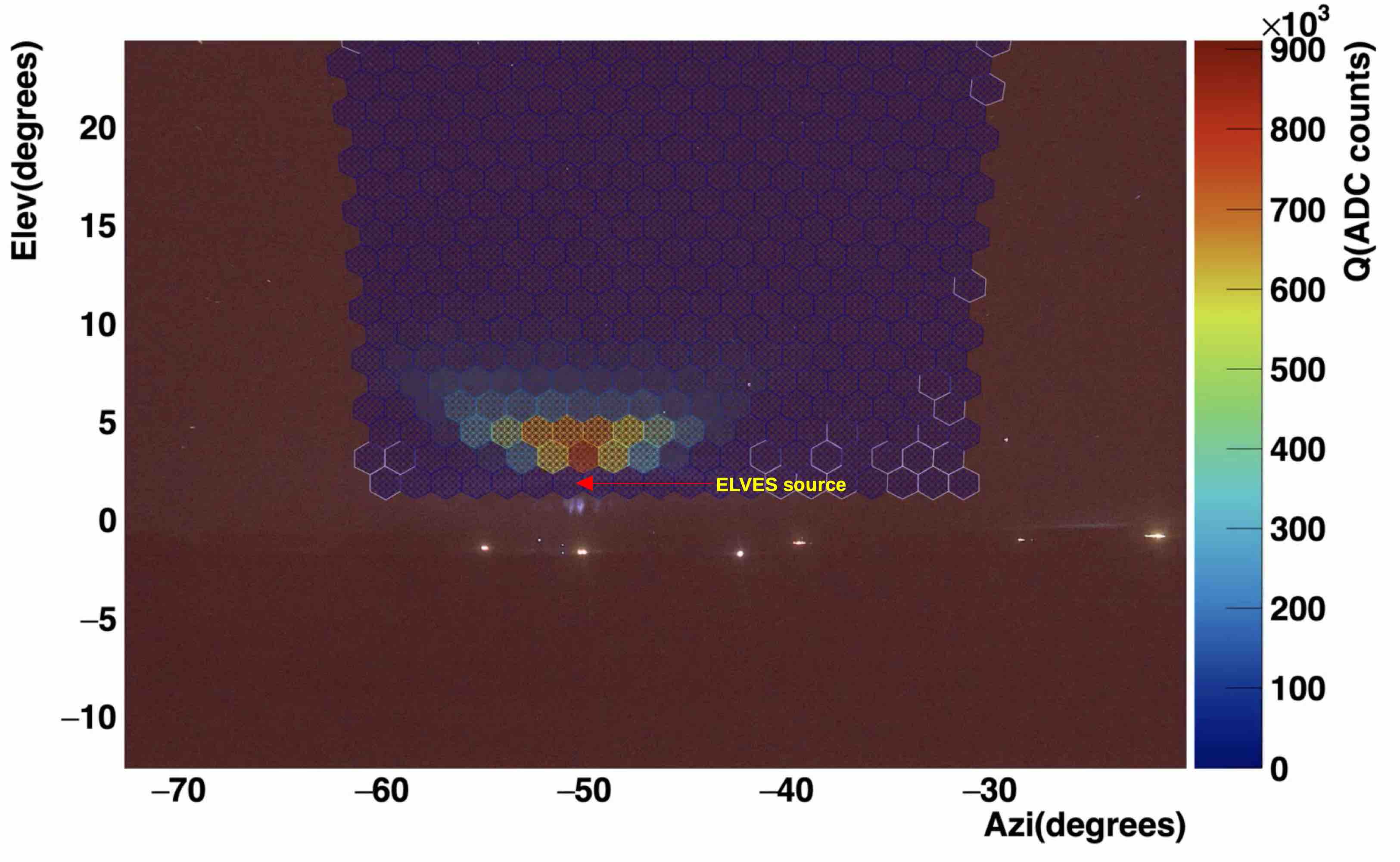} 
    \includegraphics[height=4.5cm]{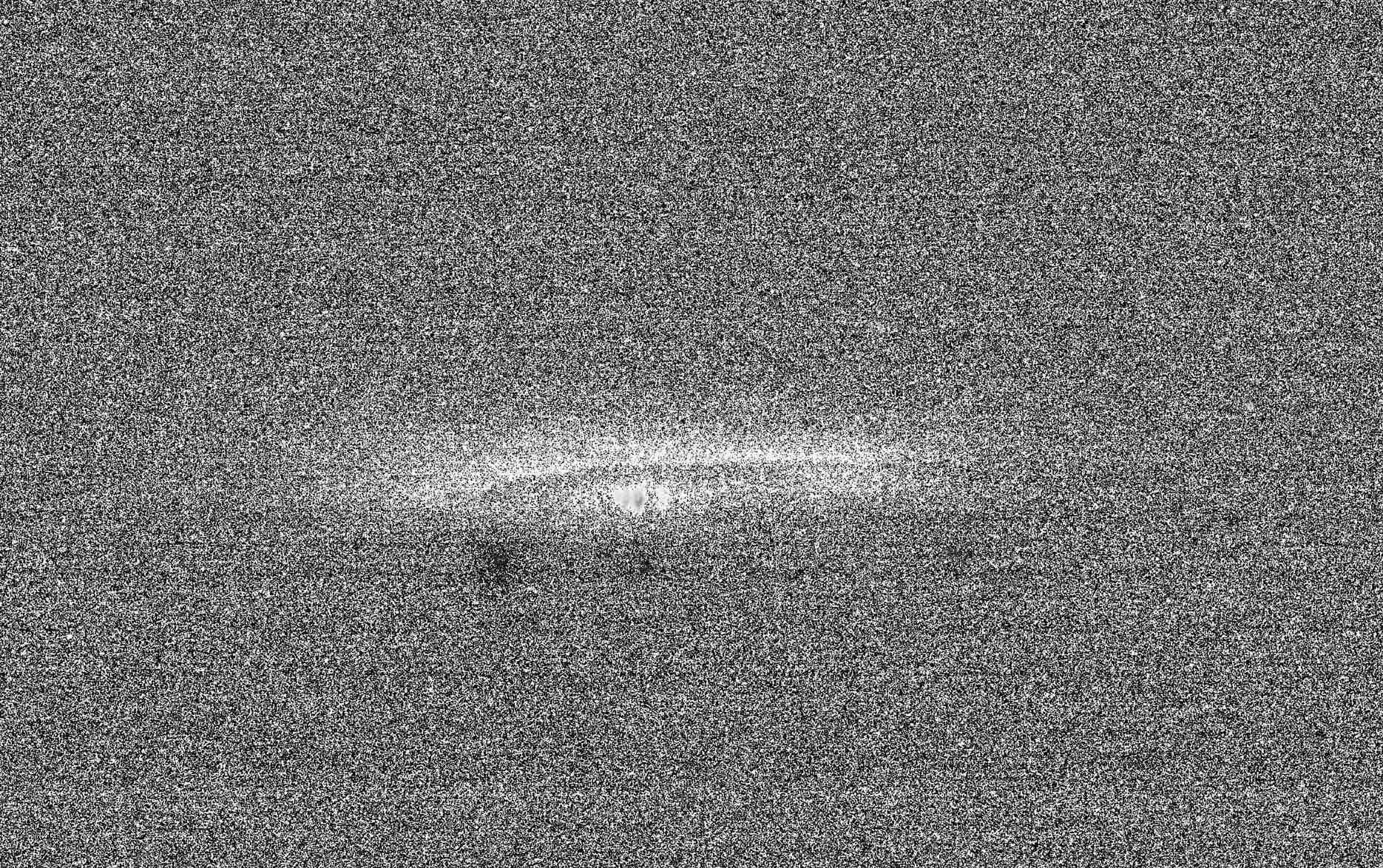}

    \caption{Left: Sprite detected at GPS time = 1416802034, overlapped on the last image recorded by the Los Leones FD, corrected for the parallax relative to Coihueco; Right: the result of the subtraction of two adjacent frames shows not just the SPRITE but also a ring of light with a diameter of about 130 km, which is interpreted as a halo. }
    \label{fig:sprite1_TLE2_LL}
\end{figure}

The second camera, TLECAM-2, was installed in April 2024, but no more storms in the field of view occurred in the following periods of data taking. Typically, the storm season in Argentina peaks in Austral spring and summer, and therefore, we had to wait for November 28, 2024, when three large storms in three hours produced up to thirty SPRITEs in the field of view of TLECAMs, but only one of these events had both an ELVES and a SPRITE in coincidence.  
Such event, observed at GPS time = 1416802034 by two Auger FD
telescopes (HEAT positioned pointing downwards \cite{Mathes:2011zz,PierreAuger:2023gmv}, i.e. between 2 and 30 degrees, and Los Leones) had several interesting properties: (a) the source was reconstructed at a quite
large distance (about 950 km), (b) is a double ELVES, (c) a ring of
light around the SPRITE is clearly visible in the TLEcam-2
image, visible in Fig.\ref{fig:sprite1_TLE2_LL}, right, obtained by subtracting two successive frames.

As we do not expect our TLE cameras to be sensitive to ELVES more than 350 km away from the FD, we conclude that the ring recorded by TLEcam-2 is
actually a halo, followed by the SPRITE. From the shape of the light curves and the analysis of the second-level trigger data (T2), which will be described in the next section, we could evince that (a) the halo and the TLE started more than 1 ms after the ELVES.
The reconstruction in Fig.\ref{fig:sprite1_TLE2_LL}, left, shows the last 2 $\mu$s recorded in HEAT, overlayed on the picture from TLEcam-2. The arrow indicates the location of the ELVES center resulting from our reconstruction.

\section{FD-TLEcam time synchronisation}

\begin{figure}[h]
\centering
    \includegraphics[width=17cm]{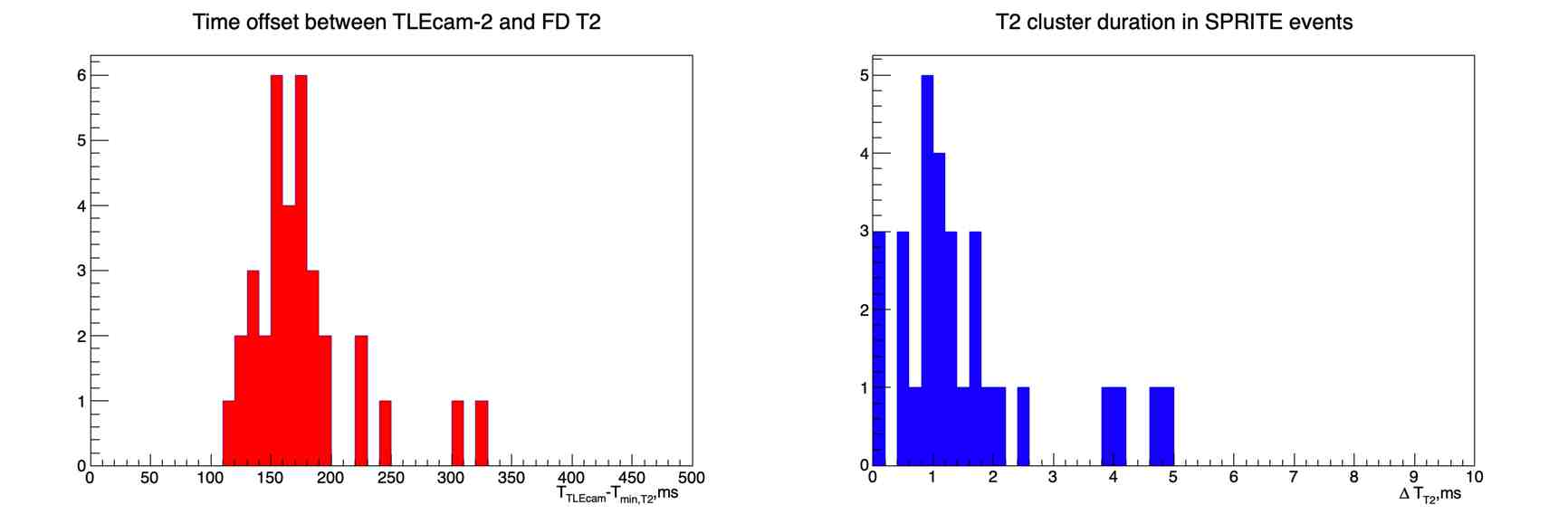} 

    \caption{Left: the difference between the time stamp of the SPRITE labeled by TLEcam-2 and the first T2 recorded by the FD. Right: the time duration of a SPRITE as measured from the T2s in the FD. }
    \label{fig:TLEcam-FD_timesync}
\end{figure}

The timing of the TLEcam images is given by the Unix time of the PC. TLEcam-1 records the starting time of each 5-minute movie, and for each subsequent frame, a recording speed of 12.5 frames per second (fps) is assumed. TLEcam-2 time tags each frame independently, recording 60 ms images with a frame rate of ~16 fps. 
The poor time resolution does not allow for a proper comparison of our
 TLE observations with other lightning detection networks, such as ENTLN \cite{Zhu:2021}. 
To improve the timing of each TLE recording, we are studying the FD timing of light pulses which pass the second level of trigger (T2) whose time stamps are saved in a local database, together with other information such as the number of triggered pixels in a given telescope, and the length of the T2 buffer waiting to be processed by the 3rd level of trigger (T3). The T2 level requires five adjacent pixels to be hit in a row-like pattern.
Typically, ELVES produce a few consecutive T2s (less than ten) with a large number of hit pixels, very often in more than one telescope. In contrast, SPRITEs can trigger a large number of consecutive T2s (up to 64 in our buffers, corresponding to 6.4 ms) with very few hit pixels, usually in just one telescope. As an example, we give the timing distributions for the events recorded on January 24, 2025, when more than sixty SPRITEs were generated by a large mesoscale convective system.
 The left plot in Fig.\ref{fig:TLEcam-FD_timesync} illustrates the time difference between the  TLEcam-2 frame nominal time and the first T2 in the cluster identified in the same GPS second. The right plot shows the duration (in ms) of the series of adjacent T2s. 

\section{Conclusions and Acknowledgments}

If the observation of ELVES at the Auger Observatory has provided more insights about the time evolution of these fast transients, one of the still open questions concerns the causal connection between different types of TLEs, which seem to originate from independent mechanisms occurring in the evolution of the lightning discharge. During this first period of commissioning, the  TLECAMs described in this paper have been manually controlled to follow the evolution of the storms. An asynchronous method for an automated offline selection of the TLEs has been designed and optimized to operate efficiently even without ELVES triggers in coincidence, but using additional time information provided from the second level of trigger in the FD.  

The realization of these new instruments would not have been possible without the invaluable help of J.~R.~Rodriguez, N.~Leal, F.~Gobbi of the Malarg\"{u}e staff, and  F.~Borotto of the INFN Torino Mechanical Workshop.

\clearpage

\section*{The Pierre Auger Collaboration}
\small

\begin{wrapfigure}[8]{l}{0.11\linewidth}
\vspace{-5mm}
\includegraphics[width=0.98\linewidth]{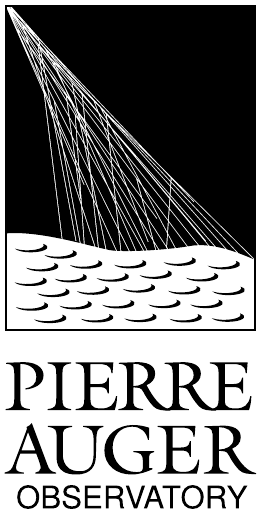}
\end{wrapfigure}

\begin{sloppypar}\noindent
A.~Abdul Halim$^{13}$,
P.~Abreu$^{72}$,
M.~Aglietta$^{54,52}$,
I.~Allekotte$^{1}$,
K.~Almeida Cheminant$^{70}$,
A.~Almela$^{7,12}$,
R.~Aloisio$^{45,46}$,
J.~Alvarez-Mu\~niz$^{79}$,
J.~Ammerman Yebra$^{79}$,
G.A.~Anastasi$^{54,52}$
% ...

\end{sloppypar}

\begin{center}
\rule{0.1\columnwidth}{0.5pt}
\raisebox{-0.4ex}{\scriptsize$\bullet$}
\rule{0.1\columnwidth}{0.5pt}
\end{center}

\vspace{-1ex}
\footnotesize
\begin{description}[labelsep=0.2em,align=right,labelwidth=0.7em,labelindent=0em,leftmargin=2em,noitemsep]
\item[$^{1}$] Centro At\'omico Bariloche and Instituto Balseiro (CNEA-UNCuyo-CONICET), San Carlos de Bariloche, Argentina
\item[$^{2}$] Departamento de F\'\i{}sica and Departamento de Ciencias de la Atm\'osfera y los Oc\'eanos, FCEyN, Universidad de Buenos Aires and CONICET, Buenos Aires, Argentina
\item[$^{3}$] IFLP, Universidad Nacional de La Plata and CONICET, La Plata, Argentina
\item[$^{4}$] Instituto de Astronom\'\i{}a y F\'\i{}sica del Espacio (IAFE, CONICET-UBA), Buenos Aires, Argentina
\item[$^{5}$] Instituto de F\'\i{}sica de Rosario (IFIR) -- CONICET/U.N.R.\ and Facultad de Ciencias Bioqu\'\i{}micas y Farmac\'euticas U.N.R., Rosario, Argentina
\item[$^{6}$] Instituto de Tecnolog\'\i{}as en Detecci\'on y Astropart\'\i{}culas (CNEA, CONICET, UNSAM), and Universidad Tecnol\'ogica Nacional -- Facultad Regional Mendoza (CONICET/CNEA), Mendoza, Argentina
\item[$^{7}$] Instituto de Tecnolog\'\i{}as en Detecci\'on y Astropart\'\i{}culas (CNEA, CONICET, UNSAM), Buenos Aires, Argentina
\item[$^{8}$] International Center of Advanced Studies and Instituto de Ciencias F\'\i{}sicas, ECyT-UNSAM and CONICET, Campus Miguelete -- San Mart\'\i{}n, Buenos Aires, Argentina
\item[$^{9}$] Laboratorio Atm\'osfera -- Departamento de Investigaciones en L\'aseres y sus Aplicaciones -- UNIDEF (CITEDEF-CONICET), Argentina
\item[$^{10}$] Observatorio Pierre Auger, Malarg\"ue, Argentina
\item[$^{11}$] Observatorio Pierre Auger and Comisi\'on Nacional de Energ\'\i{}a At\'omica, Malarg\"ue, Argentina
\item[$^{12}$] Universidad Tecnol\'ogica Nacional -- Facultad Regional Buenos Aires, Buenos Aires, Argentina
\item[$^{13}$] University of Adelaide, Adelaide, S.A., Australia
\item[$^{14}$] Universit\'e Libre de Bruxelles (ULB), Brussels, Belgium
\item[$^{15}$] Vrije Universiteit Brussels, Brussels, Belgium
\item[$^{16}$] Centro Federal de Educa\c{c}\~ao Tecnol\'ogica Celso Suckow da Fonseca, Petropolis, Brazil
\item[$^{17}$] Instituto Federal de Educa\c{c}\~ao, Ci\^encia e Tecnologia do Rio de Janeiro (IFRJ), Brazil
\item[$^{18}$] Universidade de S\~ao Paulo, Escola de Engenharia de Lorena, Lorena, SP, Brazil
\item[$^{19}$] Universidade de S\~ao Paulo, Instituto de F\'\i{}sica de S\~ao Carlos, S\~ao Carlos, SP, Brazil
\item[$^{20}$] Universidade de S\~ao Paulo, Instituto de F\'\i{}sica, S\~ao Paulo, SP, Brazil
\item[$^{21}$] Universidade Estadual de Campinas, IFGW, Campinas, SP, Brazil
\item[$^{22}$] Universidade Estadual de Feira de Santana, Feira de Santana, Brazil
\item[$^{23}$] Universidade Federal de Campina Grande, Centro de Ciencias e Tecnologia, Campina Grande, Brazil
\item[$^{24}$] Universidade Federal do ABC, Santo Andr\'e, SP, Brazil
\item[$^{25}$] Universidade Federal do Paran\'a, Setor Palotina, Palotina, Brazil
\item[$^{26}$] Universidade Federal do Rio de Janeiro, Instituto de F\'\i{}sica, Rio de Janeiro, RJ, Brazil
\item[$^{27}$] Universidade Federal do Rio de Janeiro (UFRJ), Observat\'orio do Valongo, Rio de Janeiro, RJ, Brazil
\item[$^{28}$] Universidade Federal Fluminense, EEIMVR, Volta Redonda, RJ, Brazil
\item[$^{29}$] Universidad de Medell\'\i{}n, Medell\'\i{}n, Colombia
\item[$^{30}$] Universidad Industrial de Santander, Bucaramanga, Colombia
\item[$^{31}$] Charles University, Faculty of Mathematics and Physics, Institute of Particle and Nuclear Physics, Prague, Czech Republic
\item[$^{32}$] Institute of Physics of the Czech Academy of Sciences, Prague, Czech Republic
\item[$^{33}$] Palacky University, Olomouc, Czech Republic
\item[$^{34}$] CNRS/IN2P3, IJCLab, Universit\'e Paris-Saclay, Orsay, France
\item[$^{35}$] Laboratoire de Physique Nucl\'eaire et de Hautes Energies (LPNHE), Sorbonne Universit\'e, Universit\'e de Paris, CNRS-IN2P3, Paris, France
\item[$^{36}$] Univ.\ Grenoble Alpes, CNRS, Grenoble Institute of Engineering Univ.\ Grenoble Alpes, LPSC-IN2P3, 38000 Grenoble, France
\item[$^{37}$] Universit\'e Paris-Saclay, CNRS/IN2P3, IJCLab, Orsay, France
\item[$^{38}$] Bergische Universit\"at Wuppertal, Department of Physics, Wuppertal, Germany
\item[$^{39}$] Karlsruhe Institute of Technology (KIT), Institute for Experimental Particle Physics, Karlsruhe, Germany
\item[$^{40}$] Karlsruhe Institute of Technology (KIT), Institut f\"ur Prozessdatenverarbeitung und Elektronik, Karlsruhe, Germany
\item[$^{41}$] Karlsruhe Institute of Technology (KIT), Institute for Astroparticle Physics, Karlsruhe, Germany
\item[$^{42}$] RWTH Aachen University, III.\ Physikalisches Institut A, Aachen, Germany
\item[$^{43}$] Universit\"at Hamburg, II.\ Institut f\"ur Theoretische Physik, Hamburg, Germany
\item[$^{44}$] Universit\"at Siegen, Department Physik -- Experimentelle Teilchenphysik, Siegen, Germany
\item[$^{45}$] Gran Sasso Science Institute, L'Aquila, Italy
\item[$^{46}$] INFN Laboratori Nazionali del Gran Sasso, Assergi (L'Aquila), Italy
\item[$^{47}$] INFN, Sezione di Catania, Catania, Italy
\item[$^{48}$] INFN, Sezione di Lecce, Lecce, Italy
\item[$^{49}$] INFN, Sezione di Milano, Milano, Italy
\item[$^{50}$] INFN, Sezione di Napoli, Napoli, Italy
\item[$^{51}$] INFN, Sezione di Roma ``Tor Vergata'', Roma, Italy
\item[$^{52}$] INFN, Sezione di Torino, Torino, Italy
\item[$^{53}$] Istituto di Astrofisica Spaziale e Fisica Cosmica di Palermo (INAF), Palermo, Italy
\item[$^{54}$] Osservatorio Astrofisico di Torino (INAF), Torino, Italy
\item[$^{55}$] Politecnico di Milano, Dipartimento di Scienze e Tecnologie Aerospaziali , Milano, Italy
\item[$^{56}$] Universit\`a del Salento, Dipartimento di Matematica e Fisica ``E.\ De Giorgi'', Lecce, Italy
\item[$^{57}$] Universit\`a dell'Aquila, Dipartimento di Scienze Fisiche e Chimiche, L'Aquila, Italy
\item[$^{58}$] Universit\`a di Catania, Dipartimento di Fisica e Astronomia ``Ettore Majorana``, Catania, Italy
\item[$^{59}$] Universit\`a di Milano, Dipartimento di Fisica, Milano, Italy
\item[$^{60}$] Universit\`a di Napoli ``Federico II'', Dipartimento di Fisica ``Ettore Pancini'', Napoli, Italy
\item[$^{61}$] Universit\`a di Palermo, Dipartimento di Fisica e Chimica ''E.\ Segr\`e'', Palermo, Italy
\item[$^{62}$] Universit\`a di Roma ``Tor Vergata'', Dipartimento di Fisica, Roma, Italy
\item[$^{63}$] Universit\`a Torino, Dipartimento di Fisica, Torino, Italy
\item[$^{64}$] Benem\'erita Universidad Aut\'onoma de Puebla, Puebla, M\'exico
\item[$^{65}$] Unidad Profesional Interdisciplinaria en Ingenier\'\i{}a y Tecnolog\'\i{}as Avanzadas del Instituto Polit\'ecnico Nacional (UPIITA-IPN), M\'exico, D.F., M\'exico
\item[$^{66}$] Universidad Aut\'onoma de Chiapas, Tuxtla Guti\'errez, Chiapas, M\'exico
\item[$^{67}$] Universidad Michoacana de San Nicol\'as de Hidalgo, Morelia, Michoac\'an, M\'exico
\item[$^{68}$] Universidad Nacional Aut\'onoma de M\'exico, M\'exico, D.F., M\'exico
\item[$^{69}$] Universidad Nacional de San Agustin de Arequipa, Facultad de Ciencias Naturales y Formales, Arequipa, Peru
\item[$^{70}$] Institute of Nuclear Physics PAN, Krakow, Poland
\item[$^{71}$] University of \L{}\'od\'z, Faculty of High-Energy Astrophysics,\L{}\'od\'z, Poland
\item[$^{72}$] Laborat\'orio de Instrumenta\c{c}\~ao e F\'\i{}sica Experimental de Part\'\i{}culas -- LIP and Instituto Superior T\'ecnico -- IST, Universidade de Lisboa -- UL, Lisboa, Portugal
\item[$^{73}$] ``Horia Hulubei'' National Institute for Physics and Nuclear Engineering, Bucharest-Magurele, Romania
\item[$^{74}$] Institute of Space Science, Bucharest-Magurele, Romania
\item[$^{75}$] University Politehnica of Bucharest, Bucharest, Romania
\item[$^{76}$] Center for Astrophysics and Cosmology (CAC), University of Nova Gorica, Nova Gorica, Slovenia
\item[$^{77}$] Experimental Particle Physics Department, J.\ Stefan Institute, Ljubljana, Slovenia
\item[$^{78}$] Universidad de Granada and C.A.F.P.E., Granada, Spain
\item[$^{79}$] Instituto Galego de F\'\i{}sica de Altas Enerx\'\i{}as (IGFAE), Universidade de Santiago de Compostela, Santiago de Compostela, Spain
\item[$^{80}$] IMAPP, Radboud University Nijmegen, Nijmegen, The Netherlands
\item[$^{81}$] Nationaal Instituut voor Kernfysica en Hoge Energie Fysica (NIKHEF), Science Park, Amsterdam, The Netherlands
\item[$^{82}$] Stichting Astronomisch Onderzoek in Nederland (ASTRON), Dwingeloo, The Netherlands
\item[$^{83}$] Universiteit van Amsterdam, Faculty of Science, Amsterdam, The Netherlands
\item[$^{84}$] Case Western Reserve University, Cleveland, OH, USA
\item[$^{85}$] Colorado School of Mines, Golden, CO, USA
\item[$^{86}$] Department of Physics and Astronomy, Lehman College, City University of New York, Bronx, NY, USA
\item[$^{87}$] Michigan Technological University, Houghton, MI, USA
\item[$^{88}$] New York University, New York, NY, USA
\item[$^{89}$] University of Chicago, Enrico Fermi Institute, Chicago, IL, USA
\item[$^{90}$] University of Delaware, Department of Physics and Astronomy, Bartol Research Institute, Newark, DE, USA
\item[$^{91}$] University of Wisconsin-Madison, Department of Physics and WIPAC, Madison, WI, USA
\item[] -----
\item[$^{a}$] Louisiana State University, Baton Rouge, LA, USA
\item[$^{b}$] also at University of Bucharest, Physics Department, Bucharest, Romania
\item[$^{c}$] School of Physics and Astronomy, University of Leeds, Leeds, United Kingdom
\item[$^{d}$] now at Agenzia Spaziale Italiana (ASI).\ Via del Politecnico 00133, Roma, Italy
\item[$^{e}$] Fermi National Accelerator Laboratory, Fermilab, Batavia, IL, USA
\item[$^{f}$] now at Graduate School of Science, Osaka Metropolitan University, Osaka, Japan
\item[$^{g}$] now at ECAP, Erlangen, Germany
\item[$^{h}$] Max-Planck-Institut f\"ur Radioastronomie, Bonn, Germany
\item[$^{i}$] also at Kapteyn Institute, University of Groningen, Groningen, The Netherlands
\item[$^{j}$] Colorado State University, Fort Collins, CO, USA
\item[$^{k}$] Pennsylvania State University, University Park, PA, USA
\end{description}

\vspace{-1ex}
\footnotesize
\section*{Acknowledgments}

\begin{sloppypar}
The successful installation, commissioning, and operation of the Pierre
Auger Observatory would not have been possible without the strong
commitment and effort from the technical and administrative staff in
Malarg\"ue. We are very grateful to the following agencies and
organizations for financial support:
\end{sloppypar}

\begin{sloppypar}
Argentina -- Comisi\'on Nacional de Energ\'\i{}a At\'omica; Agencia Nacional de
Promoci\'on Cient\'\i{}fica y Tecnol\'ogica (ANPCyT); Consejo Nacional de
Investigaciones Cient\'\i{}ficas y T\'ecnicas (CONICET); Gobierno de la
Provincia de Mendoza; Municipalidad de Malarg\"ue; NDM Holdings and Valle
Las Le\~nas; in gratitude for their continuing cooperation over land
access; Australia -- the Australian Research Council; Belgium -- Fonds
de la Recherche Scientifique (FNRS); Research Foundation Flanders (FWO);
Brazil -- Conselho Nacional de Desenvolvimento Cient\'\i{}fico e Tecnol\'ogico
(CNPq); Financiadora de Estudos e Projetos (FINEP); Funda\c{c}\~ao de Amparo \`a
Pesquisa do Estado de Rio de Janeiro (FAPERJ); S\~ao Paulo Research
Foundation (FAPESP) Grants No.~2019/10151-2, No.~2010/07359-6 and
No.~1999/05404-3; Minist\'erio da Ci\^encia, Tecnologia, Inova\c{c}\~oes e
Comunica\c{c}\~oes (MCTIC); Czech Republic -- Grant No.~MSMT CR LTT18004,
LM2015038, LM2018102, CZ.02.1.01/0.0/0.0/16{\textunderscore}013/0001402,
CZ.02.1.01/0.0/0.0/18{\textunderscore}046/0016010 and
CZ.02.1.01/0.0/0.0/17{\textunderscore}049/0008422; France -- Centre de Calcul
IN2P3/CNRS; Centre National de la Recherche Scientifique (CNRS); Conseil
R\'egional Ile-de-France; D\'epartement Physique Nucl\'eaire et Corpusculaire
(PNC-IN2P3/CNRS); D\'epartement Sciences de l'Univers (SDU-INSU/CNRS);
Institut Lagrange de Paris (ILP) Grant No.~LABEX ANR-10-LABX-63 within
the Investissements d'Avenir Programme Grant No.~ANR-11-IDEX-0004-02;
Germany -- Bundesministerium f\"ur Bildung und Forschung (BMBF); Deutsche
Forschungsgemeinschaft (DFG); Finanzministerium Baden-W\"urttemberg;
Helmholtz Alliance for Astroparticle Physics (HAP);
Helmholtz-Gemeinschaft Deutscher Forschungszentren (HGF); Ministerium
f\"ur Kultur und Wissenschaft des Landes Nordrhein-Westfalen; Ministerium
f\"ur Wissenschaft, Forschung und Kunst des Landes Baden-W\"urttemberg;
Italy -- Istituto Nazionale di Fisica Nucleare (INFN); Istituto
Nazionale di Astrofisica (INAF); Ministero 
dell'Universit\'a e della Ricerca (MUR); CETEMPS Center of Excellence;
Ministero degli Affari Esteri (MAE), ICSC Centro Nazionale di Ricerca in
High Performance Computing, Big Data and Quantum Computing, funded by
European Union NextGenerationEU, reference code CN{\textunderscore}00000013; M\'exico --
Consejo Nacional de Ciencia y Tecnolog\'\i{}a (CONACYT) No.~167733;
Universidad Nacional Aut\'onoma de M\'exico (UNAM); PAPIIT DGAPA-UNAM; The
Netherlands -- Ministry of Education, Culture and Science; Netherlands
Organisation for Scientific Research (NWO); Dutch national
e-infrastructure with the support of SURF Cooperative; Poland --
Ministry of Education and Science, grants
No.~DIR/WK/2018/11 and 2022/WK/12; National Science
Centre, grants No.~2016/22/M/ST9/00198,
2016/23/B/ST9/01635, 2020/39/B/ST9/01398, and 2022/45/B/ST9/02163;
Portugal -- Portuguese national funds and FEDER funds within Programa
Operacional Factores de Competitividade through Funda\c{c}\~ao para a Ci\^encia
e a Tecnologia (COMPETE); Romania -- Ministry of Research, Innovation
and Digitization, CNCS-UEFISCDI, contract no.~30N/2023 under Romanian
National Core Program LAPLAS VII, grant no.~PN 23\,21\,01\,02 and project
number PN-III-P1-1.1-TE-2021-0924/TE57/2022, within PNCDI III; Slovenia
-- Slovenian Research Agency, grants P1-0031, P1-0385, I0-0033, N1-0111;
Spain -- Ministerio de Econom\'\i{}a, Industria y Competitividad
(FPA2017-85114-P and PID2019-104676GB-C32), Xunta de Galicia (ED431C
2017/07), Junta de Andaluc\'\i{}a (SOMM17/6104/UGR, P18-FR-4314) Feder Funds,
RENATA Red Nacional Tem\'atica de Astropart\'\i{}culas (FPA2015-68783-REDT) and
Mar\'\i{}a de Maeztu Unit of Excellence (MDM-2016-0692); USA -- Department of
Energy, Contracts No.~DE-AC02-07CH11359, No.~DE-FR02-04ER41300,
No.~DE-FG02-99ER41107 and No.~DE-SC0011689; National Science Foundation,
Grant No.~0450696; The Grainger Foundation; Marie Curie-IRSES/EPLANET;
European Particle Physics Latin American Network; and UNESCO.
\end{sloppypar}

\end{document}